\documentclass[a4paper,11pt]{article}
\usepackage{jinstpub} 
\usepackage{lineno}
\usepackage{siunitx}
\usepackage{subcaption}
\usepackage{caption}
\usepackage{caption3}
\usepackage[numbers]{natbib}
\usepackage{orcidlink}

\notoc 

\title{\boldmath Performance studies of the CE-65v2 MAPS prototype structure}

\author[a,1]{A. Lorenzetti\orcidlink{0009-0009-5729-4525}\note{Corresponding author.}}
\author[a,1]{A. Ilg\orcidlink{0000-0001-9488-8095}}

\author[b]{H. Baba}
\author[c]{J. Baudot\orcidlink{0000-0001-5585-0991}}
\author[c]{A. Besson\orcidlink{0000-0001-8822-3548}}
\author[c]{S. Bugiel\orcidlink{0000-0002-0884-1440}}
\author[d]{T. Chujo\orcidlink{0000-0001-5433-969X}}
\author[c]{C. Colledani\orcidlink{0009-0001-2572-2347}}
\author[c]{A. Dorokhov\orcidlink{0000-0001-5809-524X}}
\author[c]{Z. El Bitar\orcidlink{0000-0001-6989-2703}}
\author[c]{M. Goffe\orcidlink{0000-0001-7300-4879}}
\author[b]{T. Gunji\orcidlink{0000-0002-6769-599X}}
\author[c]{C. Hu-Guo\orcidlink{0000-0001-9626-4673}}
\author[c]{K. Jaaskelainen\orcidlink{0000-0002-7172-0449}}
\author[e]{T. Katsuno\orcidlink{0009-0005-4583-6500}}
\author[f]{A. Kluge\orcidlink{0000-0002-6497-3974}}
\author[g]{A. Kostina\orcidlink{0000-0002-1356-7124}}
\author[c]{A. Kumar\orcidlink{0000-0002-4346-7335}}
\author[a]{A. Macchiolo\orcidlink{0000-0003-0199-6957}}
\author[f]{M. Mager\orcidlink{0009-0002-2291-691X}}
\author[d]{J. Park\orcidlink{0000-0002-2540-2394}}
\author[a,h]{E. Ploerer\orcidlink{0000-0001-9336-4847}}
\author[d]{S. Sakai\orcidlink{0000-0003-1380-0392}}
\author[c]{S. Senyukov\orcidlink{0000-0003-1907-9786}}
\author[c]{H. Shamas}
\author[d]{D. Shibata}
\author[f]{W. Snoeys\orcidlink{0000-0003-3541-9066}}
\author[g]{P. Stanek\orcidlink{0000-0002-6922-0461}}
\author[f]{M. Suljic\orcidlink{0000-0002-4490-1930}}
\author[g]{L. Tomasek\orcidlink{0000-0002-5224-1936}}
\author[c]{I. Valin\orcidlink{0009-0005-6517-9755}}
\author[e]{R. Wada}
\author[e]{Y. Yamaguchi\orcidlink{0009-0009-3842-7345}}

\collaboration[c]{\normalsize on behalf of the ALICE collaboration}

\affiliation[a]{Physik-Institut, University of Zurich, Winterthurerstrasse 190, 8057 Zurich, Switzerland}
\affiliation[b]{Center for Nuclear Study (CNS), University of Tokyo, 7 Chome-3-1 Hongo, Tokyo 113-8654, Japan}
\affiliation[c]{Université de Strasbourg, CNRS, IPHC UMR 7178, F-67000 Strasbourg, France}
\affiliation[d]{University of Tsukuba, 1 Chome-1-1 Tennodai, Ibaraki 305-8577, Japan}
\affiliation[e]{Physics Program, Graduate School of Advanced Science and Engineering, Hiroshima University, 1 Chome-3-2 Kagamiyama, Hiroshima 739-0046, Japan}
\affiliation[f]{European Organization for Nuclear Research (CERN), Espl. des Particules 1, 1217 Geneva, Switzerland}
\affiliation[g]{Department of Physics, Czech Technical University (CTU) in Prague, Zikova 4, 16636 Prague, Czech Republic}
\affiliation[h]{Inter-University Institute for High Energies (IIHE), Vrĳe Universiteit Brussel, Pleinlaan 2, 1050 Brussels, Belgium}

\emailAdd{alessandra.lorenzetti@cern.ch}
\emailAdd{armin.ilg@cern.ch}

\abstract{With the next upgrade of the ALICE inner tracking system (ITS3) as its primary focus, a set of small MAPS test structures have been developed in the 65 nm TPSCo CMOS process. The CE-65 focuses on the characterisation of the analogue charge collection properties of this technology. The latest iteration, the CE-65v2, was produced in different processes (standard, with a low-dose n-type blanket, and blanket with gap between pixels), pixel pitches (15, 18, \SI{22.5}{\micro\meter}), and pixel arrangements (square or staggered). The comparatively large pixel array size of $48\times24$ pixels in CE-65v2 allows the uniformity of the pixel response to be studied, among other benefits.

The CE-65v2 chip was characterised in a test beam at the CERN SPS. A first analysis showed that hit efficiencies of $\geq 99\%$ and spatial resolution better than \SI{5}{\micro\meter} can be achieved for all pitches and process variants. For the standard process, thanks to larger charge sharing, even spatial resolutions below \SI{3}{\micro\meter} are reached, in line with vertex detector requirements for the FCC-ee.

This contribution further investigates the data collected at the SPS test beam. Thanks to the large sensor size and efficient data collection, a large amount of statistics was collected, which allows for detailed in-pixel studies to see the efficiency and spatial resolution as a function of the hit position within the pixels. Again, different pitches and process variants are compared.    
}


\begin{document}
\maketitle
\flushbottom

\section{Introduction}
\label{sec:intro}

Tracking detectors are a crucial component of particle collider experiments. They detect the passage of charged particles, with the charge and momentum inferred by measuring the curvature of the reconstructed particle trajectory (\textit{track}). The precise measurement of tracks is essential to reconstruct the locations of the particle interactions and decays: the \textit{vertices}. 

The ALICE ITS3 \cite{the_technical_2024} project will replace the inner tracker of the ALICE experiment at the Large Hadron Collider (LHC) with three layers of thin, wafer-scale, bent Monolithic Active Pixel Sensors (MAPS). MAPS combine the passive sensor with the active readout chip into one silicon die. They enable a lower material budget, reduced power consumption, and smaller pitches than hybrid sensors. The sensors are manufactured in the 65 nm TPSCo CMOS process, reducing the power consumption compared to the previous generation of MAPS and enabling the construction of larger sensors given the 12-inch wafer diameter. Using the stitching technique
, a handful of curved, wafer-scale sensors is sufficient to construct three cylindrical tracking layers.


The Future Circular Collider (FCC \cite{abada_fcc-ee_2019}) is a proposed successor to the LHC. In its first phase, the FCC-ee would collide electrons and positrons at unprecedented luminosities at centre-of-mass energies between ${\sim}90$ and \SI{365}{\giga\electronvolt}. The requirements for FCC-ee vertex detectors (VXDs) are similar to those of ITS3 but demand further development in all areas, as shown in Table~\ref{tab:alice_its3_fcc_vertex}. 

\begin{table}[htbp]
    \caption{Selected design requirements of ALICE ITS3 \cite{the_technical_2024} and the FCC-ee vertex detector \cite{ilg_pixel_2024}.}
    \label{tab:alice_its3_fcc_vertex}
    \centering
    \begin{tabular}{lll}
    \textbf{Requirements}                             & \textbf{ALICE ITS3}     & \textbf{FCC-ee vertex}           \\ \hline
    Sensor spatial resolution                         & \SI{5}{\micro\meter}               & \SI{3}{\micro\meter}                        \\
    Material budget per layer [\% of $X_0$]        & 0.07 \%                 & \textless{} 0.3 \%               \\
    Radiation tolerance $[1\mathrm{MeV n_{eq}/cm^2}]$ & ${\sim}10^{13}$               & Several ${\sim}10^{13}$ per year          \\
    First layer radius            & 19 mm                   & $\lesssim \SI{13.7}{\milli\meter}$                          \\
    Power density in pixel matrix                                    & 40 $\mathrm{mW/cm^2}$   & $\lesssim 50$ $\mathrm{mW/cm^2}$ \\
    Particle hit density                               & 8.5 $\mathrm{MHz/cm^2}$ & $\mathcal{O}(200 \mathrm{MHz/cm^2})$         
    \end{tabular}
\end{table}

For the FCC-ee VXD, the first layer is envisioned to be at a radius $\lesssim \SI{13.7}{\milli\meter}$ and would potentially have to deal with a much larger particle hit rate while keeping the power density comparable. A jump in resolution is also foreseen going to only \SI{3}{\micro\meter} sensor spatial resolution. The requirement on the material budget target in terms of \% of radiation length ($X_0$) per layer is more relaxed, but FCC-ee VXDs would greatly profit from a material budget as low as ITS3 \cite{Ilg2023}.

\section{The CE-65v2 MAPS prototype structure}
\label{sec:CE-65}

The CE-65 was developed in the context of ITS3 developments. The chip combines an analogue readout with a large pixel matrix. It is used to study the charge collection properties of the 65 nm process and compare three different amplification schemes \cite{gautam_characterisation_2024}. The second version of the CE-65 prototype structure, the CE-65v2, was designed by IPHC Strasbourg and aims to go beyond the needs of ITS3. It aims to reach \SI{3}{\micro\meter} spatial resolution to meet the requirement of FCC-ee VXDs. The sensor features a large pixel matrix of $48\times24$ pixels read out by an analogue rolling shutter readout, where the matrix is read out row-by-row. The in-pixel circuitry consists of AC-coupled amplifiers, DC-separated from the input stage of the readout electronics. The reverse bias is thus not limited by the power supply for the electronics \cite{bugiel_charge_2022}. The chip comes in 15 different variants, but in the following, the focus is on pixel pitches of 15 and \SI{22.5}{\micro\meter}, both produced in the standard or modified with gap process with a classic, square, pixel arrangement. 

In the \textit{standard process} (STD), the depletion layer is balloon-shaped and does not extend to the pixel edges. This makes the charge collection diffusion-dominated with a large amount of charge sharing between neighbouring pixels and thus a good analogue spatial resolution \cite{snoeys_optimization_2023}. This process is less radiation tolerant because it is more subject to charge trapping. 

In the \textit{modified with gap process} (GAP), the pixel structure was modified to include a deep low-dose n-type blanket below the collection electrode, with a gap at the pixel edges. This increases the lateral electric field, ensuring charge collection by drift rather than diffusion to the pixel edges. This results in faster charge collection even at the edges and better radiation tolerance. The disadvantage is that lower charge sharing leads to an inferior spatial resolution \cite{snoeys_optimization_2023}.

The previous characterisation campaign focused on the overall efficiency and spatial resolution at a reverse bias of \SI{10}{\volt} \cite{ploerer_characterisation_2025}. In this contribution, reverse biases of \SI{4}{\volt} are also investigated, and the performance is evaluated depending on the hit position within the pixel (in-pixel studies).

\section{Testbeam setup and analysis}

The CE-65v2 chip was characterised in a test beam at the CERN SPS north area facility (H6 beamline \cite{cern_sps}). A mixed hadron beam of \SI{120}{GeV} was shot through the beam telescope holding the CE-65v2 device under test (DUT), six ALPIDE reference planes, and a DPTS chip also produced in \SI{65}{nm} \cite{AglieriRinella2023} for triggering\note{The authors thank the ALICE Collaboration for providing the telescope and the associated DAQ software.}. The DUT was kept at a stable temperature of \SI{20}{\degreeCelsius} using a chiller. The telescope resolution was evaluated using a telescope optimiser\footnote{\href{https://mmager.web.cern.ch/telescope/tracking.html}{https://mmager.web.cern.ch/telescope/tracking.html}} to achieve a resolution of \SI{2.2}{\micro\meter}.

Lab characterisation was performed prior to the test beam using a $^{55}\text{Fe}$ source to calibrate the sensor response \cite{ploerer_characterisation_2025}. The test beam data was then analysed using the Corryvreckan framework \cite{Dannheim:2020jlk}. In the first step, noisy pixels 
were masked both on the reference and DUT planes to avoid misalignment. Next, multiple steps of telescope alignment were performed, with one hit required on every reference plane. Lastly, the DUT is aligned with the rest of the telescope. Two methods are used to reconstruct clusters of pixels in the DUT, which are used to determine the cluster position through centre-of-gravity reconstruction: The \textit{cluster method} defines a neighbour threshold equal to the seed threshold, giving a conservative estimate of the sensor resolution. The \textit{window method} sums up all charges around the seed in a $3\times3$ window (effectively using a neighbour threshold of zero) to fully benefit from charge sharing information. Finally, the telescope resolution is subtracted in quadrature to obtain the final sensor spatial resolution. More details are given in Reference~\cite{ploerer_characterisation_2025}.

\section{Efficiency and resolution studies}

\begin{figure}[htbp]
    \centering
    \begin{subfigure}[t]{0.49\textwidth}
        \centering
        \includegraphics[width=\textwidth,trim=6.0cm 3cm 5cm 2.2cm,clip]{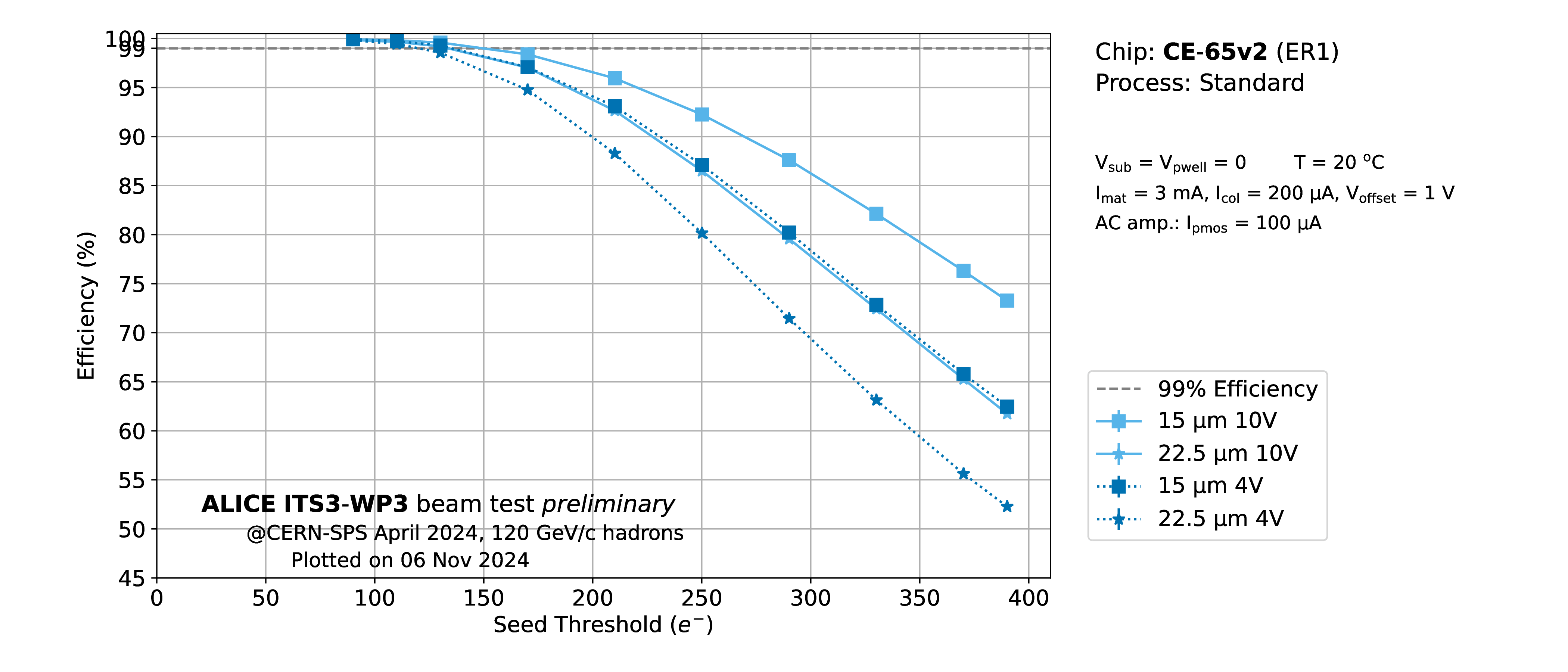}
        \caption{Efficiency, STD process.}
        \label{global_eff_STD}
    \end{subfigure}
    \hfill
    \begin{subfigure}[t]{0.49\textwidth}
        \centering
        \includegraphics[width=\textwidth,trim=6.0cm 3cm 5cm 2.2cm,clip]{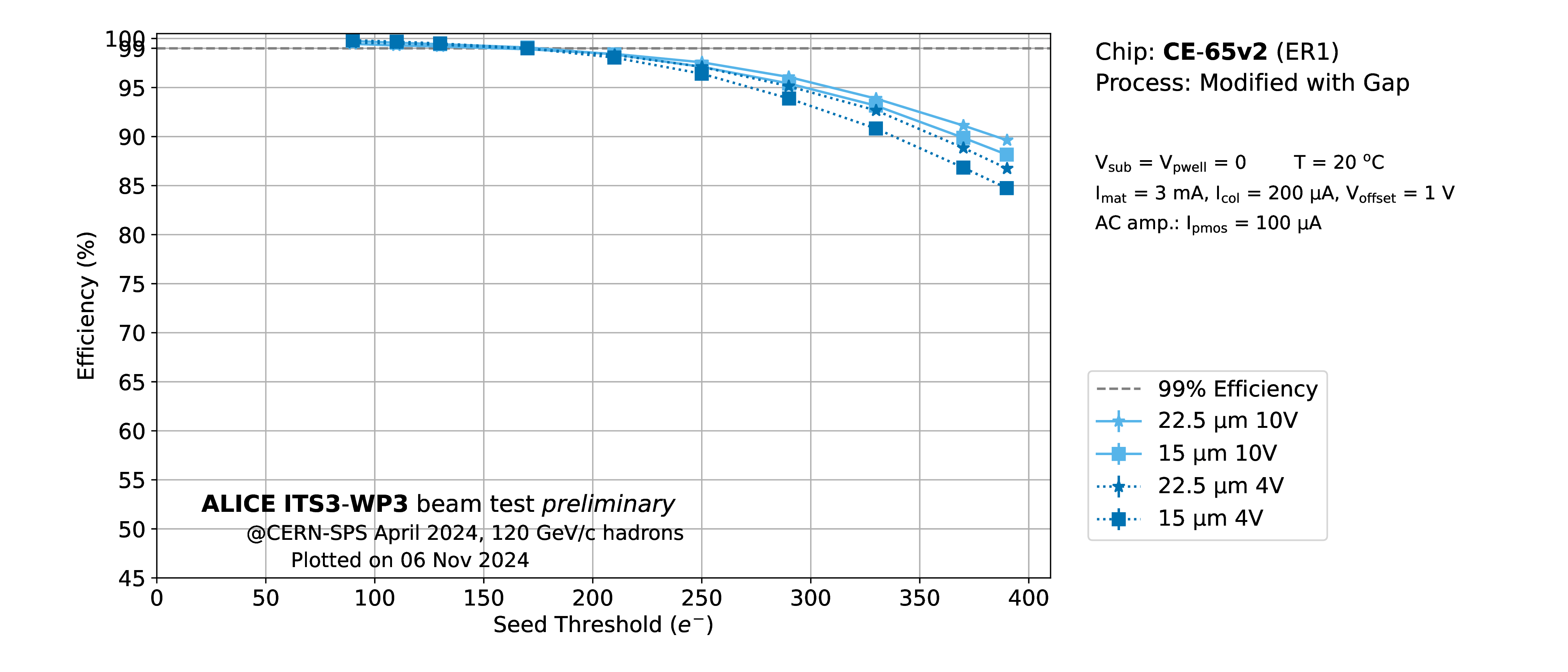}
        \caption{Efficiency, GAP process.}
        \label{global_eff_GAP}
    \end{subfigure}       
    \\
    \begin{subfigure}[b]{0.49\textwidth}
        \centering
        \includegraphics[width=\textwidth,trim=6.0cm 3cm 5cm 2.2cm,clip]{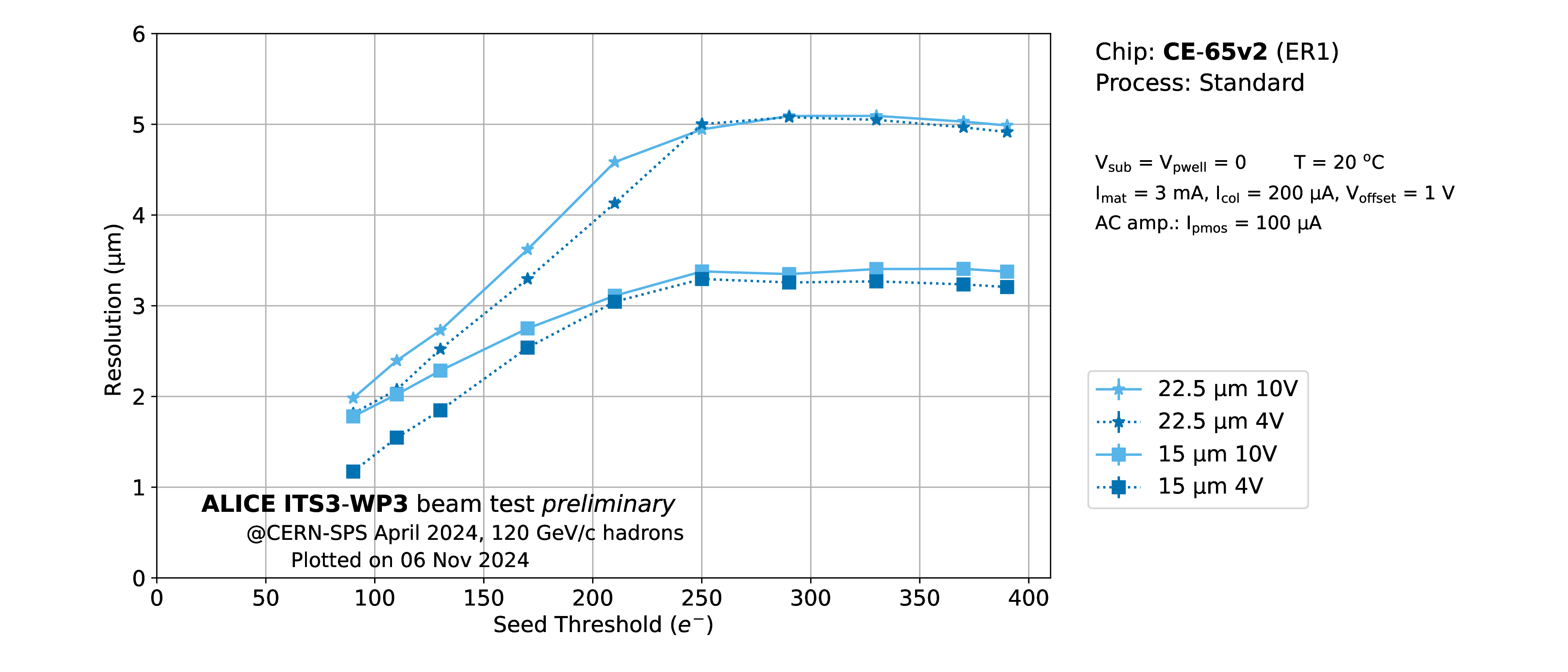}
        \caption{Spatial resolution, STD process.}
        \label{global_res_STD}
    \end{subfigure}
    \hfill
    \begin{subfigure}[b]{0.49\textwidth}
        \centering
        \includegraphics[width=\textwidth,trim=6.0cm 3cm 5cm 2.2cm,clip]{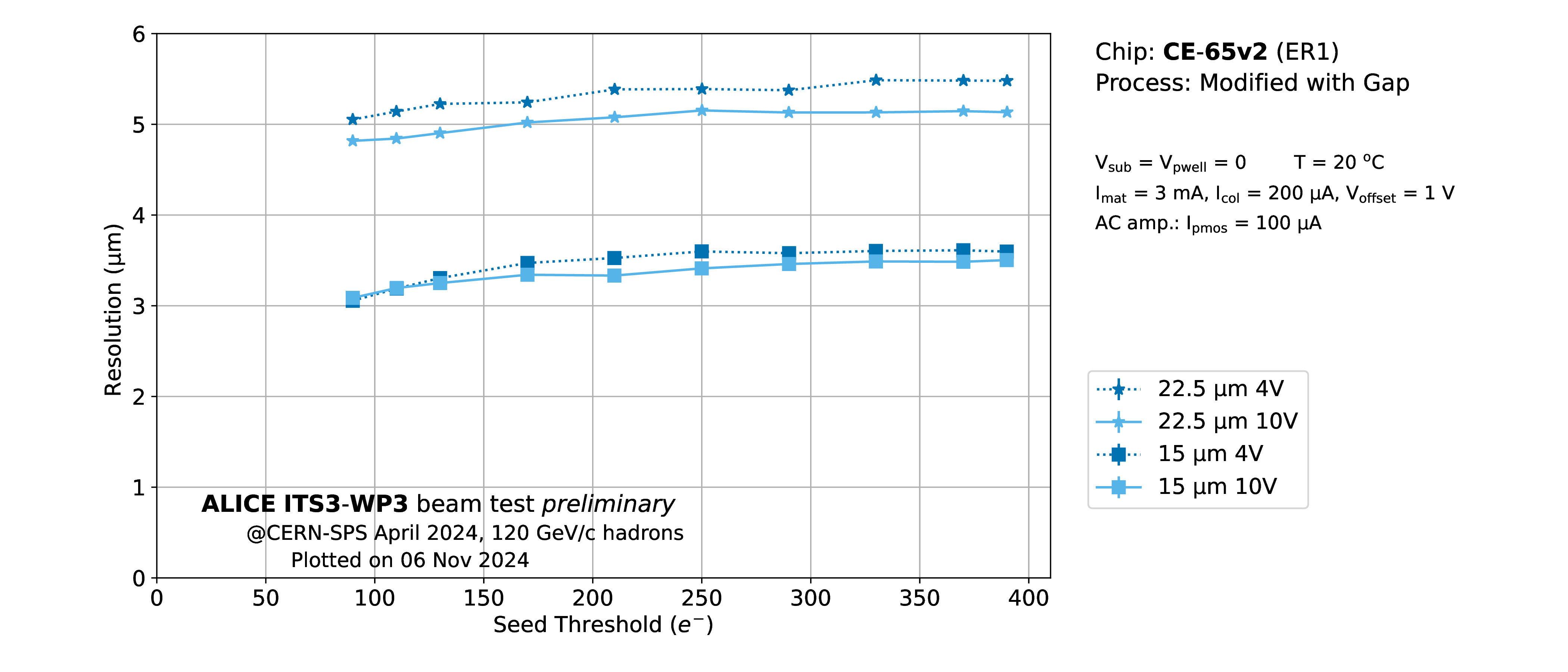}
        \caption{Spatial resolution, GAP process.}
        \label{global_res_GAP}
    \end{subfigure}    
    \caption{Efficiency (top) and spatial resolution (bottom) for different seed thresholds in the STD (left) and GAP (right) processes with pitches of 15 and \SI{22.5}{\micro\meter} and at reverse biases of 4 and \SI{10}{\volt}.}
    \label{global_eff}
\end{figure}

The efficiency and spatial resolution study uses the cluster method to reconstruct the DUT hits. Seed thresholds ranging from $90$ to $390$ $\text{e}^-$ are chosen, with the lowest threshold set equal to three times the noise RMS. The achieved global efficiencies are shown in Figs.~\ref{global_eff_STD} and \ref{global_eff_GAP}. All process, pitch, and reverse bias combinations yield efficiencies $>99\ \%$ given sufficiently low thresholds. As expected, the STD process's efficiency decreases rapidly with increasing seed threshold. In both processes, the sensor performs better at \SI{10}{\volt} reverse bias compared to \SI{4}{\volt}, indicating more depletion. For the STD process, the \SI{15}{\micro\meter} pitch performs significantly better than the larger pitch at high thresholds. This comes from the fact that the undepleted region at the pixel boundaries (where diffusion dominates) is wider for larger pitches \cite{ploerer_characterisation_2025}. This effect is not present in the GAP process, and thus, the \SI{22.5}{\micro\meter} pitch sensors feature a higher efficiency than their lower-pitch counterparts.

Figs.~\ref{global_res_STD} and \ref{global_res_GAP} show the spatial resolutions. The STD process achieves the target resolution of \SI{3}{\micro\meter} at a seed threshold of ${\sim}150\ \text{e}^-$, corresponding to efficiencies of ${\sim}99\%$ for both pitches. The GAP process does not achieve such a small resolution, but \SI{3.5}{\micro\meter} (\SI{5}{\micro\meter}) can be safely achieved at reasonable thresholds of $170\ \text{e}^-$ for a pitch of \SI{15}{\micro\meter} (\SI{22.5}{\micro\meter}), where efficiencies are $> 99\ \%$. Comparing the reverse bias voltages, it can be seen that \SI{4}{V} provides better resolution in the STD process and worse resolution in the GAP process, indicating a smaller depletion depth. In the STD process, this leads to more charge sharing and, thus, better resolution, while the GAP sensors do not benefit from this since the gap in the low-dose n-type implant inhibits charge sharing.

\section{In-pixel efficiency and mean absolute deviation studies}

The large pixel matrix of the CE-65 chip, and thus a large number of collected and reconstructed tracks, can be utilised to perform in-pixel studies. The hit information of every pixel (except for the two outermost rows/columns) is superimposed and analysed at a seed threshold of $170\ \text{e}^{-}$. For both processes, measurements of a \SI{15}{\micro\meter} pitch sensor at a reverse bias voltage of \SI{10}{V} are used. 

The \textit{cluster} method is deployed to evaluate the in-pixel efficiencies shown in Fig.~\ref{In_Pix_Eff}. In the STD process, there is a drop in performance at the edges and corners of the pixel. There, the charge is shared between multiple pixels and thus often does not overcome the seed threshold. For the GAP process, thanks to depletion up to the pixel edges and stronger lateral fields, one can observe only a minor decrease in efficiency at the corners and an otherwise flat distribution, explaining the overall better efficiency in the GAP process observed previously.

\noindent
\begin{minipage}{0.295\textwidth}
    \captionsetup{type=figure}
    \caption{In-pixel efficiency for a seed threshold of $170\ \text{e}^-$, a reverse bias voltage of \SI{10}{\volt}, and a pitch of \SI{15}{\micro\meter} in the STD (top) and the GAP process (bottom). The right side shows the evolution along a path from the pixel centre to the edge and corner and diagonally back to the centre of the pixel.}\label{In_Pix_Eff}
\end{minipage}
\begin{minipage}{0.70\textwidth}
        \includegraphics[width=\textwidth,trim=0.5cm 0.0cm 1.0cm 0.2cm,clip]{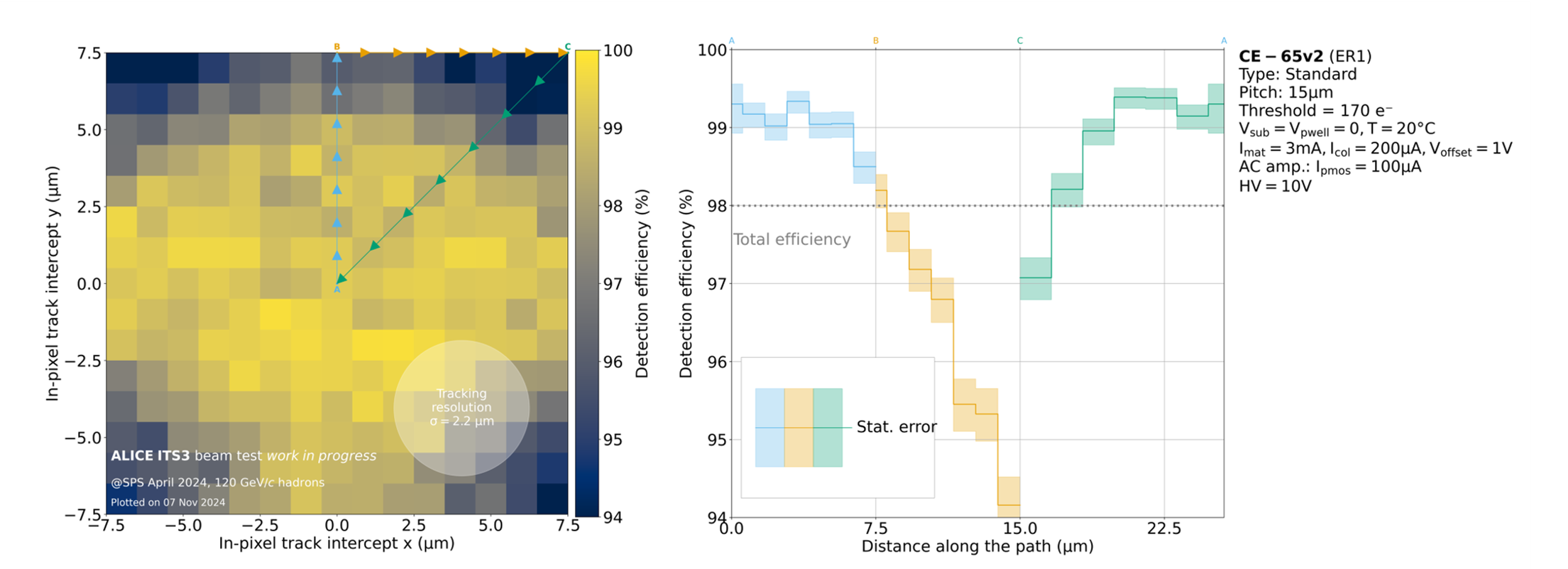}
        
        \includegraphics[width=\textwidth,trim=0.5cm 0.0cm 1.0cm 0.2cm,clip]{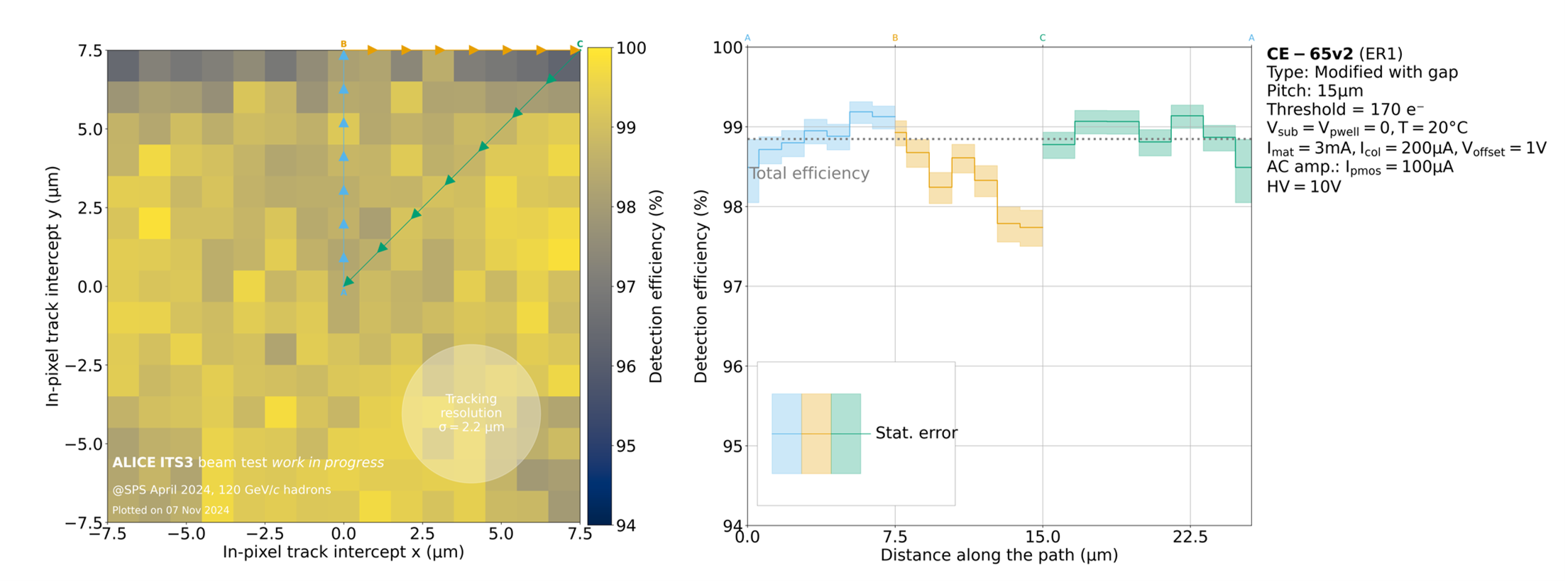}
\end{minipage}
\noindent\begin{minipage}[c]{0.70\textwidth}
    \includegraphics[width=\textwidth,trim=0.5cm 0.0cm 1.0cm 0.2cm,clip]{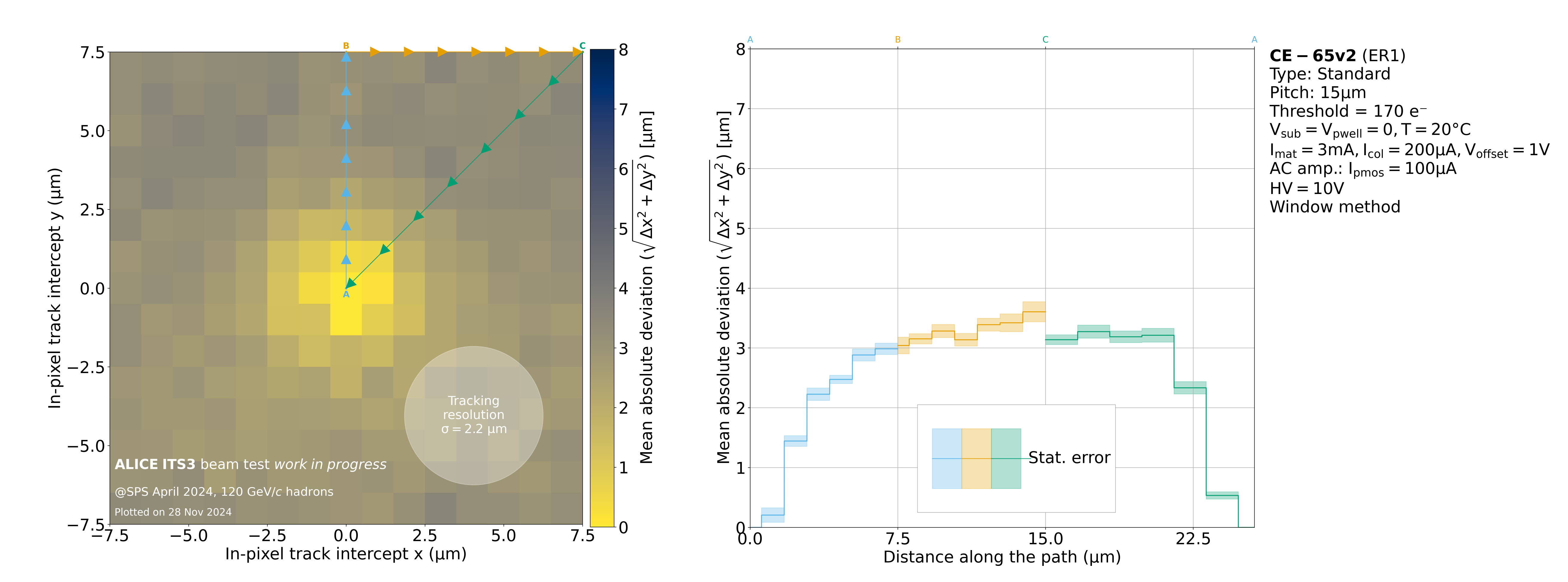}
    
    \includegraphics[width=\textwidth,trim=0.5cm 0.0cm 1.0cm 0.2cm,clip]{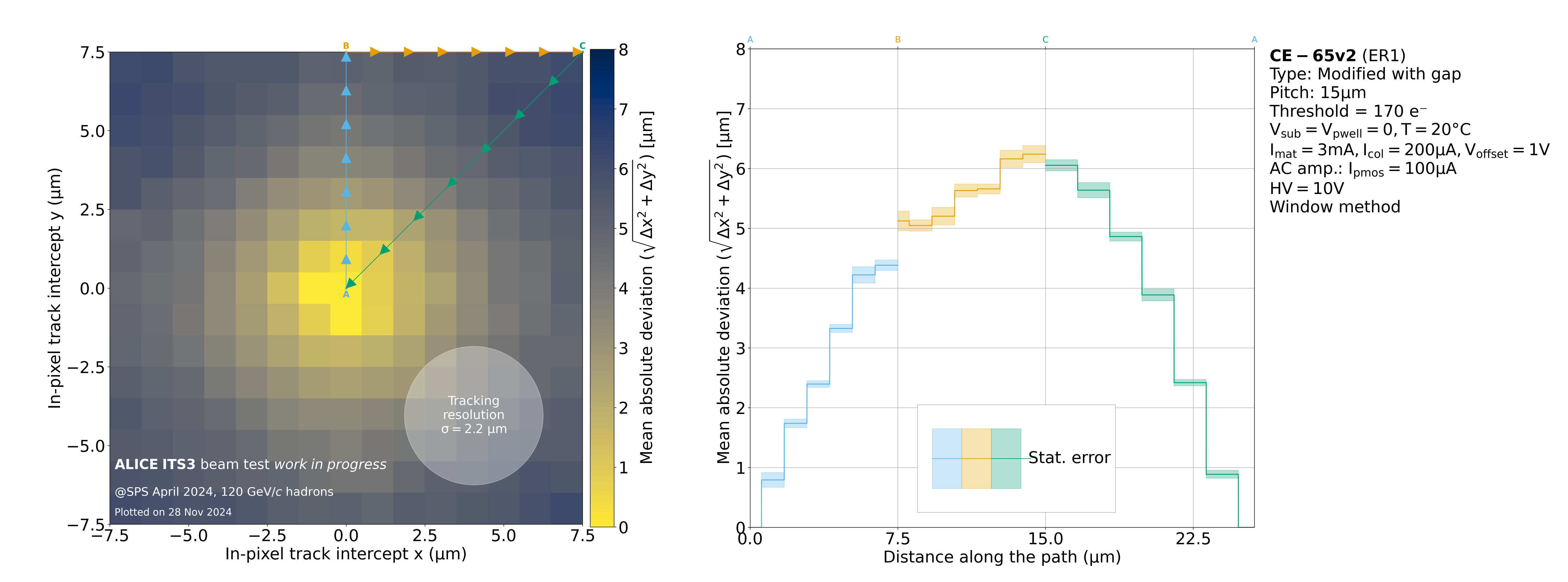}
\end{minipage}
\begin{minipage}[c]{0.295\textwidth}
    \captionsetup{type=figure}
    \caption{Mean absolute deviation for a seed threshold of $170\ \text{e}^-$, reverse bias voltage of \SI{10}{\volt}, and a pitch of \SI{15}{\micro\meter} in the STD (top) and GAP process (bottom). The right side shows the evolution along a path from the pixel centre to the edge and corner and diagonally back to the centre of the pixel.}\label{In_Pix_Res}
\end{minipage}

The in-pixel measurement of the mean absolute deviation between the reconstructed telescope and DUT cluster positions ($\sqrt{\Delta x^2 + \Delta y^2}$) uses the window method. Fig.~\ref{In_Pix_Res} shows the mean absolute deviation depending on the position of the hit inside the pixel. In the STD process, the distribution is more uniform and does not exceed $\approx \SI{3.5}{\micro\meter}$ anywhere in the pixel. For the GAP sensor, the performance strongly deteriorates at the edges and corners, which leads to the overall worse spatial resolution observed before. Using the window method, the full power of the STD process is showcased: Charge sharing from the seed to the neighbouring pixels strongly helps to reconstruct the hit position and reach resolutions much below $\text{pitch}/\sqrt{12}$.

\section{Conclusions and Outlook}

This work builds on previous measurements of the CE-65v2 \cite{ploerer_characterisation_2025} and investigates operation at reverse bias voltages of \SI{4}{\volt} and evaluated the performance depending on the in-pixel hit position. Two different paths towards superior spatial resolution in CE-65 structures were explored. Using the STD process, resolutions $<\SI{2}{\micro\meter}$ ($<\SI{3}{\micro\meter}$) are achieved with a pitch of \SI{15}{\micro\meter} (\SI{22.5}{\micro\meter}), while still operating with over $99\ \%$ efficiency. The GAP process, featuring faster charge collection but less charge sharing, enables resolutions of ${\sim}\SI{3.3}{\micro\meter}$. A further reduction of the pitch could also allow the GAP process to fulfill the FCC-ee VXD spatial resolution requirement.  

The challenges of achieving \SI{3}{\micro\meter} resolution in a large digital read-out chip efficient at reasonable noise levels and in a radiation environment differ between the two processes. In STD, an intricate analogue-to-digital conversion scheme is necessary to fully benefit from the charge sharing information while keeping the readout data rate minimal. For the GAP process, the challenge is to integrate all the necessary logic into pixels of small pitch required to achieve \SI{3}{\micro\meter} resolution.

The next steps in characterising the CE-65 are to measure the effect of irradiation and test \SI{18}{\micro\meter} pitch sensors and the Modified without Gap process. Lastly, the impact of a staggered pixel arrangement, mimicking hexagonal pixels, is also still being evaluated.

CE-65 marks a promising step towards 
MAPS fit for future lepton collider vertex detectors.

\bibliographystyle{JHEP}
\bibliography{Bibliography.bib}









\end{document}